\documentclass{aa}  

\usepackage{graphicx}
\usepackage[colorlinks=true,linkcolor=blue,citecolor=blue]{hyperref}
\usepackage{url}
\begin{document} 
\defcitealias{gallegocano2024}{IBGC--I}

   \title{Age and metallicity of the Milky Way's nuclear star cluster studied at 3\,pc from Sagittarius\,A*}

   \author{E. Gallego-Cano
          \inst{1}
         \and
         R. Sch\"odel
         \inst{1}
         \and
           T. K. Fritz
          \inst{2,3}
          \and      
      M. W. Hosek Jr.
         \inst{4}
         \and
         K. Muzic
         \inst{5}
         \and         
        A. Feldmeier-Krause
          \inst{6}
            \and
        \'A.~Mart\'inez-Arranz 
          \inst{1}
         \and
         F. Nogueras-Lara
         \inst{1}
   }
   \institute{
       Instituto de Astrof\'isica de Andaluc\'ia (CSIC),
     Glorieta de la Astronom\'ia s/n, 18008 Granada, Spain; \email{lgc@iaa.es}
     \and
     School of Data Science, University of Virginia, 1919 Ivy Road, Charlottesville, VA 22903, USA
    \and
     Fraunhofer IOSB (Institute of Optronics, System Technologies and Image Exploitation), Department Object Recognition, Gutleuthausstr. 1, 76275 Ettlingen, Germany
     \and
    Physics and Astronomy Department, University of California, Los Angeles, CA 90095-1547, USA
     \and
     Instituto de Astrof\'isica e Ci\^{e}ncias do Espa\c{c}o, Faculdade de Ci\^{e}ncias, Universidade de Lisboa, Ed. C8, Campo Grande, 1749-016 Lisbon, Portugal
     \and
    Department of Astrophysics, University of Vienna, T\"urkenschanzstraße 17, Wien, 1180, Austria
     }

   \date{Received July 15, 2025; accepted July 16, 2025}


 \abstract
  {The Milky Way’s nuclear star cluster (NSC) offers a unique laboratory for studying the formation and evolution of dense stellar systems around a supermassive black hole. Previous work suggests that most stars in the NSC are old; however, the detailed age and metallicity distributions remain uncertain.}
  {We aim to constrain the star formation history (SFH) and metallicity of a poorly explored region of the NSC located about 3 pc from Sagittarius\,A*.}
  {We analysed NACO/VLT observations in the intermediate-band filter centred at 2.24\,\(\mu\)m, complemented by $H$-band imaging, and constructed colour–magnitude diagrams and completeness-corrected $K$-band luminosity functions. We clearly identified the red clump and red giant branch bumps. The SFH was derived by fitting cumulative luminosity functions with theoretical models from MIST, PARSEC, and BaSTI, which spans a wide range of ages and metallicities, and by employing Monte Carlo sampling to estimate uncertainties. We also constrained the metallicity of the stellar population, further refined by spectroscopic data from the literature.}
  {The NSC stellar population is predominantly old and metal-rich, with $75.6 \pm9.5$~\% of the stellar mass formed $\gtrsim 10$ Gyr ago and a median metallicity of [M/H] $\sim+0.35$. Significant contributions come from an intermediate-age population around 2-3\,Gyr ($20.8 \pm8.7$~\%), while minor components appear at $\sim400$ Myr ($0.9 \pm0.8$~\%) and 20 Myr ($3.6 \pm1.4$~\%), the latter representing a small but non-negligible young population. Systematic uncertainties related to stellar evolution models, binning, photometric range, unresolved binaries, and filter selection are considered.}
  {Our findings indicate that the NSC formed predominantly in an early episode, with a substantial contribution from a star formation episode 2–3\,Gyr ago and minor younger components. The metallicity estimates are consistent with spectroscopic measurements, and the results agree with the stellar population properties of the inner NSC and the inner nuclear stellar disc, providing useful constraints on the transition between these two structures.}


  \titlerunning{Age and metallicity of the MWNSC at 3pc from Sgr\,A*}
   \maketitle

%

\section{Introduction}

Located at about $\sim8$ kpc from Earth, the Galactic centre (GC) hosts the $\sim4\times 10^{6}$\,M$_{\odot}$  black hole Sagittarius\,A*  \citep[Sgr\,A*; e.g.][]{do2019relativistic,abuter2021improved}, which is surrounded by the nuclear star cluster (NSC) with a mass of $\sim2.5 \times 10^{7}$\,M$_{\odot}$ and a half-light radius of $\sim4-5$\,pc \citep{schodel2014surface, fritz2016mass, gallegocano2020}. The proximity of the GC enables us to resolve individual stars on milliparsec scales, thus making it a unique laboratory for studying a galactic nucleus.

Observations of the NSC are challenging due to strong and locally highly variable interstellar extinction, strong source crowding, and small-scale extinction variations \cite[see e.g.][]{Schodel:2014bn}. Observations are typically restricted to wavelengths $\gtrsim1.5\,\mu$m, where 
interstellar extinction becomes less extreme and require significantly sub-arcsecond angular resolution.

Intermediate-band (IB) photometry has proven to be effective for identifying young and intermediate-age stars in the NSC \cite[e.g.][]{buchholz2009composition,nishiyama2013young, nishiyama2023search, gallegocano2024}. This technique offers the advantage of probing the spectral energy distributions of fainter sources over larger fields than is feasible with spectroscopy. \cite{gallegocano2024} (hereafter \citetalias{gallegocano2024}) analysed the central parsec, studying the distribution and initial mass function (IMF) of young, massive stars. In this paper, we focus on a region of the NSC at $\sim3$ pc from Sgr\,A* and derive new constraints on its formation history using IB photometry.

Studies of the star formation history (SFH) of the NSC have mostly concentrated on the central parsec  \cite[e.g.][]{pfuhl2011, schoedel2020sfh,chen2023}. The region 1–3\,pc from Sgr\,A* has remained poorly explored \citep[but see][]{Blum:2003fk}. This zone lies outside the small field near Sgr\,A* that is frequently observed with high-resolution. However, it lies well inside the effective radius of the NSC and may therefore be essential to understand the formation and evolution of the NSC. A key diagnostic in SFH studies of the old population can be the red giant branch bump (RGBB) \citep[e.g.][]{nataf2011} and its distance from the red clump (RC) \citep{Girardi2016} in luminosity functions (LFs), providing insights into aspects such as mean metallicity and age \citep[e.g.][]{nogueraslara2018rgbb}.

In this work, we study a region of the NSC beyond the central parsec that has not been previously examined. We use observations with the adaptive optics assisted near-infrared (NIR) camera NACO \citep{Lenzen2003naco,Rousset2003naco} at the ESO VLT\footnote{This work is based on observations made with ESO Telescopes at the Paranal Observatory under programs 087.B-0182(A), 091.B-0172(A).}. These data enabled us to detect the RC and RGBB in the target field and separate them clearly. We analyse the stellar LF to derive new insights into the SFH of the NSC.

\section{Data reduction and analysis}\label{sec:data}

\begin{figure*}
    \centering
    \includegraphics[width=\textwidth]{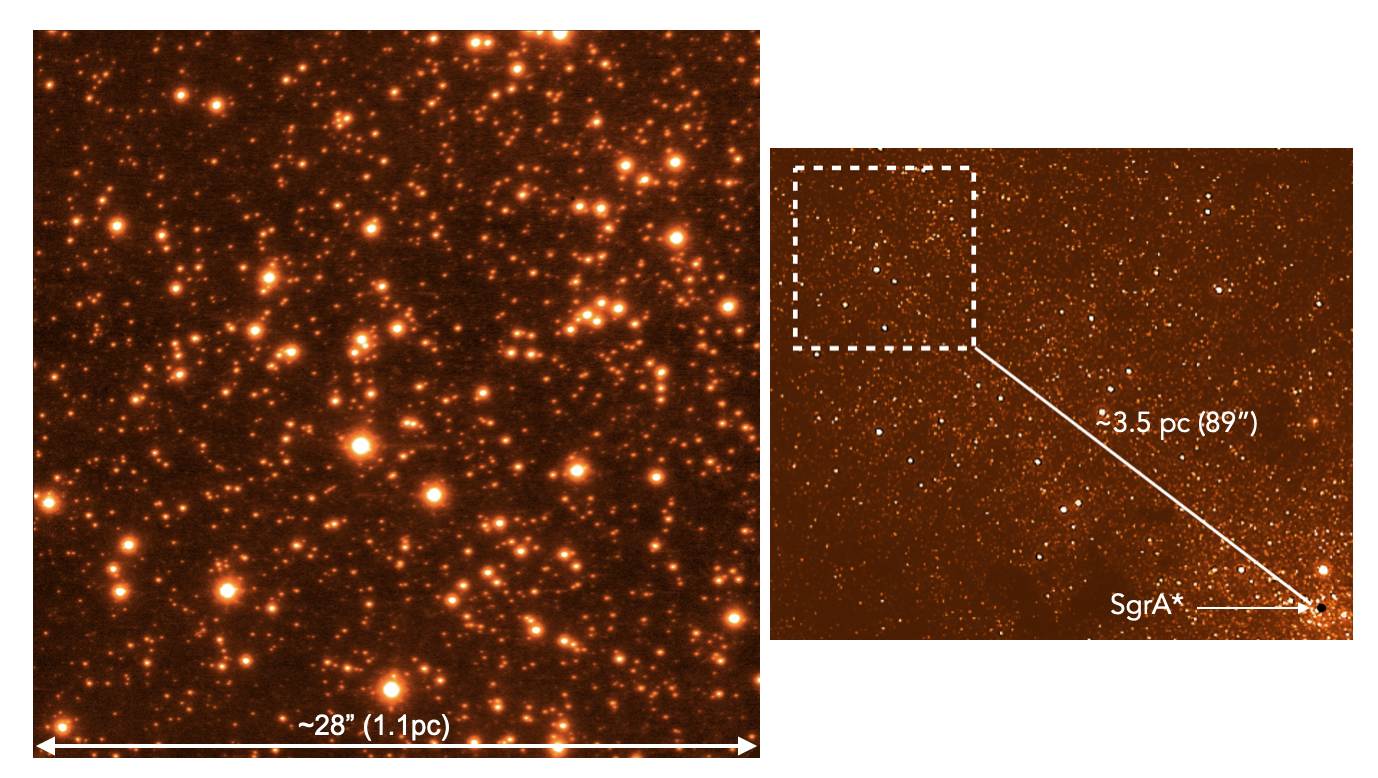}
    \caption{Left: Target field with a FoV of approximately $28'' \times 28''$. Right: Location over a $K_{S}$-band image from the GNS survey. The field lies at a projected distance of approximately $89''$ (3.5\,pc) north-east of Sgr~A*, marked by the black point.}
    \label{fig:location_field}
\end{figure*}

We used data obtained on 31 May 2011 with the S27 camera of NACO/VLT through the IB\_2.24 filter ($\lambda_{\rm central}=2.24\,\mu\mathrm{m}$, $\Delta\lambda=0.06\,\mu\mathrm{m}$). The data were taken without dithering, using the same pointing for five exposures of 19.2\,s each, resulting in a total integration time of 57.6\,s. The pixel scale is $0.027''$\,pixel$^{-1}$, and the resulting spatial resolution is approximately $0.07''$–$0.10''$. We refer to this filter as IB224. Figure~\ref{fig:location_field} shows the combined NACO image of the field in this filter (left panel).

Although other observations of the same field exist with different IB filters, we selected the IB224 filter for our analysis because it is not affected by the CO absorption band. In contrast, we avoided the IB230 and IB236 filters ($\lambda_{\rm central}=2.30$–$2.36\,\mu\mathrm{m}$, respectively) because they coincide with the CO band-head absorption of giant stars, which would lead to a reduced signal-to-noise ratio. The field of view (FoV) is approximately $28'' \times 28''$ (left panel of Fig.~\ref{fig:location_field}). The centre of the field is located at a projected distance of approximately 3.5\,pc ($89''$) north-east of Sgr~A* (see right panel of Fig.~\ref{fig:location_field}).

In addition, we used $H$-band observations of the same field obtained on 1 May 2013, consisting of six exposures with a total detector integration time of 66\,s, to identify and remove foreground and background stars. To our knowledge, neither dataset has previously been analysed.

For data reduction and analysis, we followed the procedure described in \citetalias{gallegocano2024}, with some adjustments. We applied sky subtraction, bad-pixel removal, and flat-fielding. We re-binned the data by a factor of two using quadratic interpolation to improve the astrometric and photometric quality of the final products \citep[see][]{gallegocano2018,schoedel2018}. As in \citetalias{gallegocano2024}, saturated stars were repaired to ensure accurate point-spread function (PSF) estimation, a critical step for both astrometry and photometry.

For the analysis, we incorporated noise maps, adopted a spatially variable PSF, and determined robust photometric uncertainties through bootstrapping, including PSF-related errors (see \citetalias{gallegocano2024} for further details). Astrometric calibration was performed using $K_{S}$-band data from the GALACTICNUCLEUS survey (GNS; \citealt{nogueraslara2019gns2}). 

The photometric calibration differs from \citetalias{gallegocano2024}, which used OB stars as reference sources: the number of known OB stars in the field analysed here is too small. We therefore first determined approximate zero points using GNS broadband photometry, then applied a local calibration to correct for differential extinction and homogenize the calibration quality across the FoV. We used the same RC template based on spectroscopically confirmed RC stars from the best-quality inner field as in \citetalias{gallegocano2024}. Specifically, for each target star we selected RC stars within a fixed radius of $\sim5''$ and averaged their magnitudes; the difference between this mean magnitude and the IB224 template value from the central field was used to derive the photometric correction. The adopted radius accounts for regions where extinction is strong and spatially variable. Only RC stars with photometric uncertainties smaller than 0.06 mag were included in this procedure. Stars with fewer than 20 nearby RC neighbours were flagged and excluded from the calibrated sample. 

The magnitudes obtained from the calibration procedure are internally consistent but were not originally in the Vega system. To transform them into the Vega system, we applied a wavelength-dependent extinction correction relative to the $K$ band, adopting an extinction power-law index of $-2.3$ and a reference $A_{K}=2.24$ from the central $10''$ extinction map of \citet{nogueraslara2018gns}, relative to which we calibrated the IB magnitudes (using the RC template mentioned above). 

\begin{figure} [ht!]
\includegraphics[width=\columnwidth]{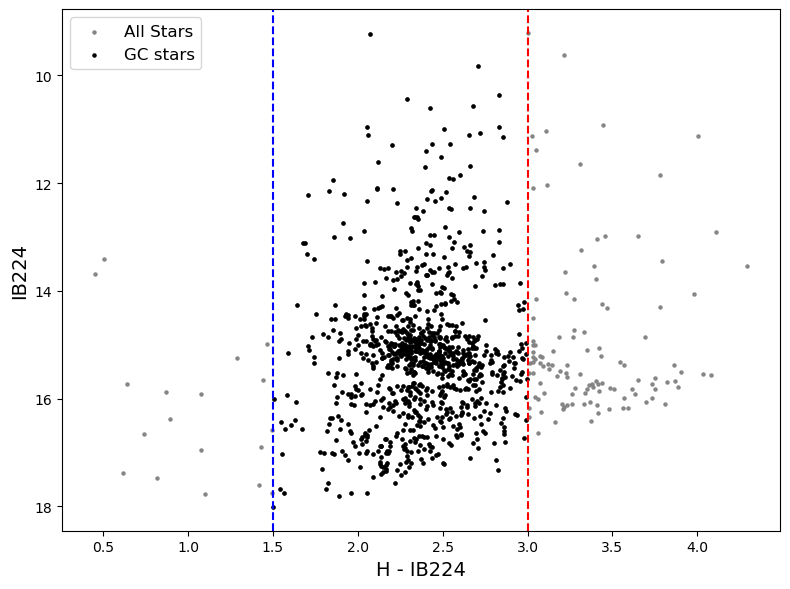}
\caption{CMD for the observed field. The blue and red lines mark the limits used to exclude foreground and background stars. The final sample of GC stars is shown in black.}
  \label{Fig:cmd_field6}
\end{figure}

Figure \ref{Fig:cmd_field6} shows the colour-magnitude diagram (CMD) for the field corrected for differential reddening. Following \citet{gallegocano2018}, we excluded stars with $H - \mathrm{IB224} < 1.5$, which we classify as foreground stars, and those with $H - \mathrm{IB224} > 3.0$, which are likely background stars or intrinsically reddened objects. A caveat is that not all stars have $H$-band counterparts; however, since we limited our analysis to IB224$ \lesssim17.2$, the incompleteness of this selection is negligible.

The completeness was estimated using artificial star tests. For each 0.5 mag step between $12.0-19.0$, we created an artificial star image by inserting 200 stars into the real image. Artificial stars were placed at random positions, but avoiding regions within $\sim0.08''$ of any real detected star to prevent artificially increasing the crowding. Each artificial star image was then processed independently using the same photometric pipeline and parameters as applied to the real data. The recovery fraction as a function of input magnitude provides the completeness curve. Based on these results, we limited the analysis to IB224 $\lesssim 17.2$, where completeness remains above 90\%. Within this magnitude range, the completeness correction has a negligible effect on the derived LF and SFH.

The final LF corrected for completeness and differential extinction, is shown by the data points and black line in Figure~\ref{Fig:klf_metallicityfits}. The final sample contains 2223 stars.

\section{Star formation history} \label{sec:SFH}

To derive the SFH, we followed the procedure described in \cite{schoedel2023nsd}, in which the observed LF is modelled as a best-fit combination of theoretical LFs. We assumed a linear combination of 15 single-age stellar populations: 0.02, 0.05, 0.1, 0.4, 1, 1.5, 2, 3, 4, 5, 6, 7, 9, 11, and 13 Gyr. The age grid is finer at younger epochs, where the LFs evolve more rapidly. We also tested a 5 Myr component, but it consistently received zero weight. This indicates that the <0.1 Gyr bins are already sufficient and that the data do not reveal a very young population.

For constructing theoretical LFs using MIST isochrones, we used \texttt{SPISEA} \citep{hosek2020spisea}, which incorporates NACO IB filters. In addition, we constructed theoretical LFs based on BaSTI \citep{hidalgobasti2018, pietrinfernibasti2021} and PARSEC \citep{bressanparsec2012, chenparsec2014, tangparsec2014, marigoparsec2017, pastorelliparsec2020} isochrones. We adopted a Salpeter IMF for BaSTI and a Kroupa IMF for MIST and PARSEC. The PARSEC LFs include corrections for unresolved binaries. \texttt{SPISEA} similarly accounts for unresolved multiplicity, while BaSTI models do not consider multiplicity. To evaluate the impact of unresolved binaries on our results, we also computed MIST-based models without multiplicity. We found no significant impact on our results. Different IMFs were also tested to assess their impact; however, no significant effect was observed since the sample is dominated by giant stars with masses of 1–2 M\textsubscript{\(\odot\)}  \citep{schoedel2020sfh}. We assumed a distance of 8.2\,kpc and created LFs for metallicities spanning [M/H] = -0.30 to +0.45 (see Appendix \ref{metal_unc}), since metallicity significantly influences the relative amplitudes and separation of the RC and RGBB \citep[see Fig.\,6 in][]{schoedel2023nsd}.

We performed 100 Monte Carlo (MC) realizations, in which the observed LF was resampled within its measurement errors, and refitted the model for each synthetic dataset. For each MC realization, we considered ten independent random initializations of the fit parameters and retained the solution with the lowest $\chi^2$. This approach mitigates the possibility of the fit converging to a local minimum. We used the cumulative LF to avoid binning effects. Uncertainties in the observed LF were estimated through bootstrap resampling. 

The free parameters of our model included the weights of the different stellar populations, the overall extinction, and a Gaussian smoothing parameter. The latter is necessary to account for effects such as photometric measurement errors, residual variations due to imperfect differential extinction correction, and the physical depth of the stellar population along the line of sight -- all of which result in a smoothing of the observed LF. The fit was restricted to $12.5-17.2$\,mag to avoid saturation at the bright end and maintain completeness at the faint end. We explored the effect of different magnitude cuts to assess the related systematic uncertainties (see Appendix~\ref{range_unc}).

\begin{figure}[ht!]
\includegraphics[width=\columnwidth]{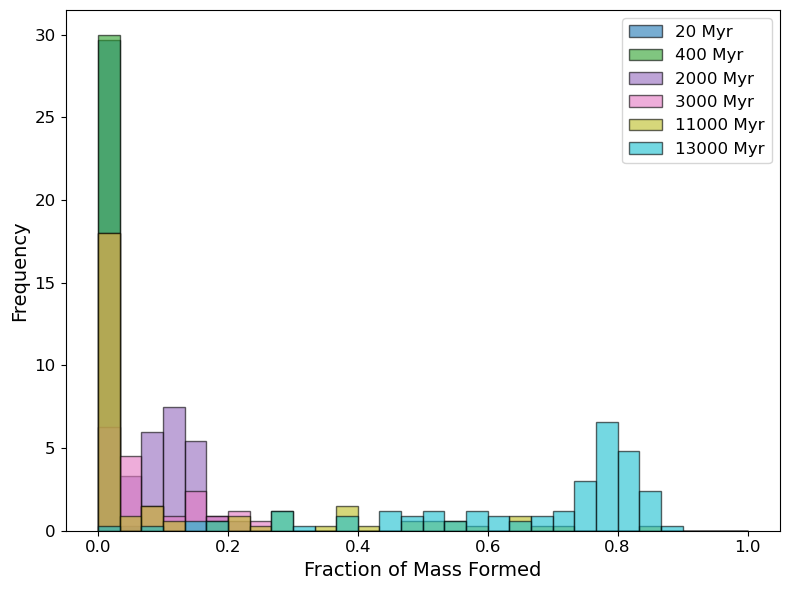}
\caption{Distribution of the best-fit stellar ages for MIST and [M/H] = +0.30, derived using 15 age bins. Only ages contributing more than 1\% of the originally formed stellar mass are shown.} 
\label{Fig:sfh_dist}
\end{figure}

Figure~\ref{Fig:sfh_dist} shows the frequency histogram of the best-fit age weights from one of our MC runs, using MIST isochrones and [M/H] = +0.30. It appears that most stars formed $> 10$ Gyr ago, with a significant contribution at 2-3\,Gyr as well as minor contributions at 400 Myr and 20 Myr. All other ages bins contributed $<1\%$ of the stellar mass formed. 
We obtained consistent results for different metallicities and for the PARSEC and BaSTI models (Appendix~\ref{metal_unc}).  

As shown in the figure, only a small number of age bins is necessary to fit the LF. To avoid overfitting and to use the minimum number of age bins necessary, we evaluated the statistical preference of different age bins using the Bayesian information criterion (BIC) \citep{Schwarz1978bic} and the stability of the recovered weights across MC realizations (Appendix~\ref{age_bin}). The combination of the four ages -- 0.02, 0.4, 2, or 3 (depending on the theoretical model used) -- and 13\,Gyr provided the lowest BIC values and the most stable results, while still reproducing all significant features of the LF satisfactorily and preventing the fit from assigning artificial weights to poorly constrained age intervals. Further details, including uncertainties from the fitting range and metallicity, are given in Appendix~\ref{cumul_analysis}.

Figure~\ref{Fig:sfh} shows the resulting SFH for the three stellar evolution models. The horizontal error bars indicate the width of the age intervals covered by each fitted component.

\begin{figure}[ht!]
\centering
\includegraphics[width=\columnwidth]{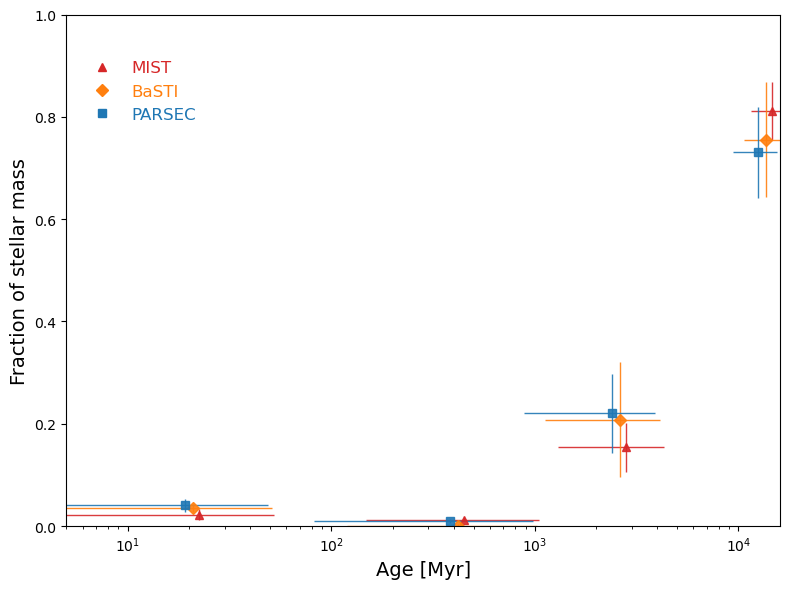} 
\caption{SFHs from the MIST, PARSEC, and BaSTI models. For clarity, the points are slightly jittered along the horizontal axis to improve visualization.}
\label{Fig:sfh}
\end{figure}
The uncertainties include statistical and systematic contributions (see Appendix~\ref{cumul_analysis}) added in quadrature. As mentioned above, although we explored different metallicities, only the models with [M/H] $\ge 0$ were used when computing the systematic uncertainty to avoid overestimating the errors (see Appendix~\ref{metal_unc}). The resulting stellar mass fractions and total uncertainties are 
$3.6 \pm 1.4$\% ($20^{+30}_{-20}~\mathrm{Myr})$, 
$0.9 \pm 0.8$\% ($400^{+600}_{-300}~\mathrm{Myr})$, 
$20.8 \pm 8.7$\% ($2.5^{+1.5}_{-1.5}~\mathrm{Gyr})$, and
$75.6 \pm 9.5$\% ($13^{+3}_{-3}~\mathrm{Gyr})$.

\section{Metallicity} \label{sec:metallicity}
The position and brightness of the RGBB are highly sensitive to metallicity: more metal-rich populations exhibit a fainter and more pronounced RGBB \citep[e.g.][]{cho2002,nogueraslara2018rgbb}. In the previous section, we explored a broad range of metallicities to quantify their impact on the derived SFH (see Appendix~\ref{metal_unc}), but only the solar and super-solar models were used for the systematic uncertainty term. In this section, we specifically examine which metallicity provides the best overall fit. In contrast to the cumulative LF used in the SFH analysis, we use the binned LF here because it allows us to measure the impact of the fit around the RGBB more directly. Following the method described above, each model LF was generated as a linear combination of 15 single-age stellar populations. In this step, we did not group ages into bins, since we are interested only in global fit quality rather than a detailed SFH. For each metallicity, we report the reduced $\chi^2$ over the full IB224 range, which quantifies the global fit quality, and the mean normalized squared residual within $15 < $IB224$ < 17$, which evaluates the local fit around the RGBB. This local statistic averages the squared residuals, scaled by their uncertainties, within this interval and offers a practical measure of model performance in this region.

Table~\ref{tab:metallicity_all} summarizes the results for all stellar evolution models and metallicities. The metallicity corresponding to the lowest reduced $\chi^2$ identifies the best global fit (marked with an asterisk, $\ast$), whereas the lowest mean residual marks the best local fit (marked with a dagger, $\dagger$). All models favour super-solar metallicities, typically between +0.3 and +0.45~dex. MIST yields the best-fitting and most locally accurate solution at [M/H] = +0.45; PARSEC favours [Fe/H] = +0.48; and BaSTI supports a mildly super-solar composition around [Fe/H] = +0.30. The $\alpha$-enhanced solar-metallicity BaSTI model provides a significantly poorer fit, indicating that $\alpha$-enhancement does not improve the match to the observed RGBB or overall LF shape. As mentioned in Section~\ref{sec:SFH}, only MIST and PARSEC include unresolved multiplicity and unresolved binaries, respectively. To confirm that this does not affect the metallicity analysis, we repeated the MIST fit at [M/H] = +0.30 without multiplicity. The resulting reduced $\chi^2$ and local residual (1.214 and 0.497) differ only marginally from the values obtained with multiplicity (1.190 and 0.486). This small change is within the uncertainties and does not influence the metallicity preference. 

\begin{figure} [ht!]
\includegraphics[width=\columnwidth]{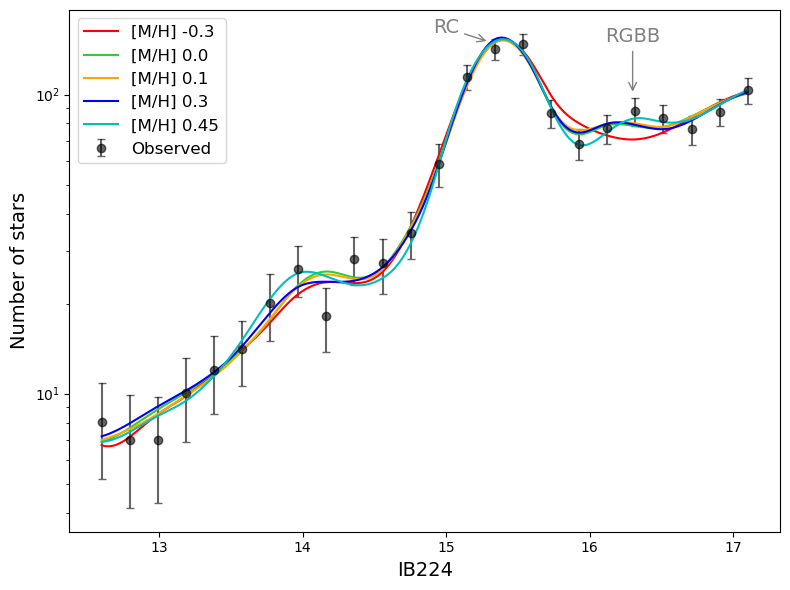}
\caption{Best fits of the theoretical LFs with varying metallicity to the de-reddened and completeness-corrected LF of the target field. We mark the RC and RGBB.}
  \label{Fig:klf_metallicityfits}
\end{figure}

Combining the preferred metallicities of the three stellar evolution models, we adopted a characteristic value of $\langle \mathrm{[M/H]} \rangle = +0.41^{+0.07}_{-0.11}$, where the central value is the mean of the best-fit model metallicities and the asymmetric uncertainties reflect the extrema of the model range (+0.48 and +0.30). This captures the systematic spread among models. Figure~\ref{Fig:klf_metallicityfits} shows the LF fits for the MIST models in the tested metallicities, illustrating the sensitivity of the RGBB feature and the overall morphology of the LF to metallicity variations. We also note the presence of the asymptotic giant branch bump at IB224 $ \sim $ 14, whose strength and position depend on metallicity and the modelling of late evolutionary phases.

\begin{table}[ht]
\centering
\small
\caption{Metallicity fits across all models for the field.}
\label{tab:metallicity_all}
\begin{tabular}{lccc}
\hline
Model & [Fe/H] or [M/H] & Reduced $\chi^2$ & Local residual\\
\hline
MIST & --0.30 & 1.975 & 0.933 \\
& 0.00 & 1.347 & 0.479 \\
& +0.10 & 1.398 & 0.537 \\
& +0.30 & 1.190 & 0.486 \\
& +0.45 & 0.899$^{\ast}$ & 0.214$^{\dagger}$  \\
\hline
PARSEC & --0.30 & 2.533 & 1.277 \\
& +0.01 & 2.232 & 1.100 \\
& +0.15 & 1.985 & 0.903 \\
& +0.30 & 1.616 & 0.683 \\
& +0.48 & 1.389$^{\ast}$ & 0.462$^{\dagger}$ \\
\hline
BaSTI & +0.06 & 1.597 & 0.895 \\
& +0.30 & 1.244$^{\ast}$ & 0.612$^{\dagger}$ \\
& +0.06 $\alpha$-enh. & 3.975 & 3.796 \\
\hline
\end{tabular}
\tablefoot{
The asterisk ($\ast$) denotes the model yielding the minimum reduced $\chi^2$ (best global fit), 
while the dagger ($\dagger$) denotes the model yielding the minimum mean normalized squared residual in the $15 < K_s < 17$ range (best local fit).}
\end{table}

We also analysed the stars from \citet{feldmeier2020asymetry} and \citet{feldmeier2025} within the field. The median metallicity of this sample (about 50 stars) is $\mathrm{[M/H]}_{\rm spec} = +0.32^{+0.07}_{-0.06}$, with uncertainties estimated via bootstrap resampling. Combining this spectroscopic measurement with our photometric metallicity results yields a combined metallicity of $\langle \mathrm{[M/H]} \rangle_{\rm combined} = +0.35^{+0.07}_{-0.06}$, consistent with both methods and confirming a characteristic super-solar metallicity for the field.

\section{Discussion and conclusions} \label{sec:discussion}

We presented a study of the SFH of the Milky Way's NSC in a field at a projected distance of about 3.5 \,pc from Sgr\,A*. The SFH was inferred from fitting theoretical LFs to the observed LFs. Using different stellar evolution models allowed us to quantify model-dependent effects and to improve the robustness of the derived ages and metallicities \citep[see][]{Morales2025}. The increasing differences between models at ages $\leq$ 1 Gyr reflect the growing sensitivity of the LF to the model-dependent assumptions in the evolved stellar phases, leading to systematic uncertainties in the inferred mass fractions.

We find an old, metal-rich population with 75.6\% of the stellar mass having formed $>10$\,Gyr ago and a median metallicity of [M/H] $\approx$ 0.35. An intermediate-age population of 2-3\,Gyr contributes $\sim21\%$ of the originally formed mass. Minor contributions are detected at $\sim400$\,Myr (0.9\%) and 20\,Myr (3.6\%). While most known young and intermediate age stars in the NSC are concentrated near Sgr~A* \citep[see e.g.][]{Genzel:2010fk,feldmeier2015kmos}, the detection of a small young component in our field suggests that recent star formation may not be strictly confined to the innermost region \citep[see also][]{Nishiyama:2016zr}.

Regarding potential contamination of the studied NSC field by stars from the nuclear stellar disc, we consider this unlikely to be significant. The colour cut may have removed some contamination; in any case, we do not expect more than $\sim$20\% of NSD stars in our field \citep[see Fig.\,5 in][]{Feldmeier-Krause:2025yx}. The inner nuclear disc has a prominent 1\,Gyr component in its SFH \citep{nogueraslara2020}, which we do not find in the LF analysed here. 

Our results are in good agreement with \cite{schoedel2020sfh}, who found that $\sim80$\% of the stars in the central parsec of the NSC formed $>$10\,Gyr ago and $\sim$15\% at $\sim3$\,Gyr.
In contrast, \citet{chen2023} found a younger dominant age of $\sim5$ Gyr in the central 1.5 pc. Their study is based on a fundamentally different methodology using spectroscopy and Bayesian forward modelling. This discrepancy is noteworthy and will need to be examined in future studies. One possible source of this discrepancy is that our LF is dominated by faint stars  ($K\gtrsim15$) in the RC and RGBB, while theirs is dominated by significantly brighter stars ($K\lesssim14-15.5$). Reassuringly, both studies find a high mean metallicity. Finally, we note that the age of the intermediate age population identified here and in \citet{schoedel2020sfh} is close to the age of the old population identified in \citet{chen2023}.

We confirm that NSC stars have super-solar mean metallicities, in agreement with other studies \citep{feldmeier2020asymetry, schoedel2020sfh, Nogueras-Lara:2022by, feldmeier2025}. Our LF fitting yields a single metallicity value for the entire population, which should be understood as the average metallicity of the light-dominating stars rather than that of every individual star. Spectroscopic work shows that a small fraction ($\sim$10\%) of stars in the NSC are metal-poor \citep{pfuhl2011, do15metallicy, Feldmeier-Krause:2017kq, gallegocano2024}, and our method is not sensitive to such a minority component. The super-solar metallicity may also imply a reduced production of neutron stars, potentially alleviating the so-called missing-pulsar problem in the GC, as discussed by \citet{chen2023}.

Our results indicate that the NSC experienced an early formation phase, followed by a significant episode around 2–3\,Gyr ago. The presence of a small but non-negligible young component further reveals star formation outside the central parsec in more recent times. This finding motivates a more detailed investigation of these regions, which we plan to carry out in future studies.

\section*{Data availability}

The table of stellar positions and photometry used in this work is available in electronic form at the CDS via anonymous ftp to cdsarc.u-strasbg.fr (130.79.128.5) or via \url{http://cdsweb.u-strasbg.fr/cgi-bin/qcat?J/A+A/}.

\begin{acknowledgements}
EGC, RS, AMA, and FNL acknowledge financial support from the Severo Ochoa grant CEX2021-001131-S funded by MCIN/AEI/ 10.13039/501100011033 and from grants PID2022-136640NB-C21 and PID2024-162148NA-I00 funded by MCIN/AEI 10.13039/501100011033 and by the European Union. AFK acknowledges funding from the Austrian Science Fund (FWF) [grant DOI 10.55776/ESP542]. K.M. acknowledges support from the Funda\c{c}\~{a}o para a Ci\^{e}ncia e a Tecnologia (FCT) through the CEEC-individual contract 2022.03809.CEECIND and research grant UID/04434/2023. FNL acknowledges financial support from the Ram\'on y Cajal programme (RYC2023-044924-I) funded by MCIN/AEI/10.13039/501100011033 and FSE+.

\end{acknowledgements}

%
%

\bibliographystyle{aa} 
\bibliography{bibliography.bib} 

\begin{appendix}

\section{Uncertainties in the fit of the cumulative function}\label{cumul_analysis}

\subsection{Age bin analysis}\label{age_bin}

We tested the robustness of the recovered SFH by performing fits using 4, 6, 9, and 13 age bins across three different stellar evolution models (MIST, PARSEC, BaSTI). All schemes provided comparable reduced $\chi^2$ values in the range $1.01-1.09$, indicating that the overall quality of the fits does not strongly depend on the number of age bins. 

To further assess model preference, we computed the BIC for each configuration. We defined the BIC as 
\begin{equation}
\mathrm{BIC} = n \cdot \ln(\chi^2_\nu) + k \cdot \ln(n),
\label{eq:BIC}
\end{equation}
where $n$ is the number of data points, $k$ is the number of free parameters, and $\chi^2_\nu$ is the reduced chi-squared of the fit. The BIC penalizes models with more parameters while favouring better fits, so lower values indicate a statistically preferred model. In all models, the solutions with 4 age bins consistently yielded the lowest BIC values, suggesting that more complex partitions were not statistically favoured. Furthermore, the 4-bin solutions also produced the most stable SFH estimates, with smaller scatter across MC realizations.

Based on the minimal $\chi^2$, low BIC, and stable SFH uncertainties, we adopted the 4-bin solution for the main analysis. The choice of age binning introduced a systematic uncertainty in the stellar mass fractions of each age bin that we incorporated in Equation~\ref{eq:unc_propagation}. For example, for the BaSTI model, we obtained $\sigma_{\rm binning}=0.0055$, $0.0049$, $0.0825$, and $0.0792$, respectively.

\subsection{Uncertainties from the fitting range}\label{range_unc}
We also tested the sensitivity to the adopted fitting range. Specifically, we repeated the analysis with small shifts in the faint limit (17.2, 17.0, 16.75) and the bright limit (12.5, 12.2), each with 100 MC realizations. The resulting median mass fractions varied only slightly across these configurations. The scatter of the medians provided an estimate of the systematic uncertainty associated with the fitting range. For example, for the MIST model we obtained $\sigma_{\rm range}=0.0065$, $0.0003$, $0.0248$, and $0.030$, respectively. These values were small compared to the statistical errors from the MC simulations, but we nevertheless included them in the final error budget.

\subsection{Uncertainties from the fitting metallicity}\label{metal_unc}
We also explored the effect of the adopted metallicity on the recovered SFH. For the MIST, we tested [M/H] values of $+0.45$, $+0.30$, $+0.10$, $0.00$, and $-0.30$, obtaining $\sigma_{\rm metal}=0.004$, $0.004$, $0.040$, and $0.041$ in the four age bins, respectively. For PARSEC, we performed an analogous test using [Fe/H] = $+0.48$, $+0.30$, $+0.10$, $0.00$, and $-0.30$, and found similar uncertainties. To avoid overestimating the errors, only metallicities $\ge 0$ were included in the metallicity contribution.

The total systematic uncertainties reported in the main analysis combine all contributions:
\begin{equation}
\sigma_{\rm final}^2
= \sigma_{\rm stat}^2
+ \sigma_{\rm binning}^2
+ \sigma_{\rm range}^2
+ \sigma_{\rm metal}^2
+ \sigma_{\rm model}^2,
\label{eq:unc_propagation}
\end{equation}
where the systematic terms are assumed to be symmetric and added in quadrature.

\end{appendix}

\end{document}